\documentclass[pre,twocolumn,aps,superscriptaddress]{revtex4}
\usepackage[T1]{fontenc}
\usepackage[latin9]{inputenc}
\usepackage{array}
\usepackage{booktabs}
\usepackage{mathrsfs}
\usepackage{mathbbol}
\usepackage{multirow}
\usepackage{amsmath}
\usepackage{amsthm}
\usepackage{amssymb}
\usepackage{stmaryrd}
\usepackage{graphicx}
\usepackage{latexsym}
\usepackage{textcomp}
\usepackage{mathtools}
\usepackage{xcolor}
\usepackage[citecolor=magenta,colorlinks=true]{hyperref}
\usepackage{lineno}
\usepackage{stackrel}
\usepackage[toc,page,titletoc]{appendix}
\usepackage{bm}
\definecolor{mycolor1}{rgb}{0.1, 0.6, 0.6}
\begin{document}

\title{Universal stress correlations in crystalline and amorphous packings}
\author{Roshan Maharana}
\email{roshanm@tifrh.res.in}
\affiliation{Tata Institute of Fundamental Research, Hyderabad 500107, India}
\author{Debankur Das}
\email{debankur.das@uni-goettingen.de}
\affiliation{
Institute for Theoretical Physics, Georg-August-Universit\"{a}t G\"{o}ttingen, 37 077 G\"{o}ttingen, Germany}
\author{Pinaki Chaudhuri}
\email{pinakic@imsc.res.in}
\affiliation{The Institute of Mathematical Sciences, Taramani, Chennai 600113, India}
\author{Kabir Ramola}
\email{kramola@tifrh.res.in}
\affiliation{Tata Institute of Fundamental Research, Hyderabad 500107, India}

\date{\today}

\begin{abstract}
We present a universal characterization of stress correlations in athermal systems, across crystalline to amorphous packings. 
Via numerical analysis of static configurations of particles interacting through harmonic as well as Lennard-Jones potentials, for a variety of preparation protocols and ranges of microscopic disorder, we show that the properties of the stress correlations at large lengthscales are surprisingly universal across all situations, independent of structural correlations, or the correlations in orientational order. In the near-crystalline limit, we present exact results for the stress correlations for both models, which work surprisingly well at large lengthscales, even in the amorphous phase.
Finally, we study the differences in stress fluctuations across the amorphization transition, where stress correlations reveal the loss of periodicity in the structure at short lengthscales with increasing disorder.
\end{abstract}

\maketitle

\section{Introduction}
Athermal solids form due to the macroscopic rigidity of their constituent particle networks, and are stable to mechanical perturbations~\cite{baule2018edwards,henkes2007entropy,bi2015statistical,cates1998jamming,tong2015crystals}. Such collective elasticity emerges in any system of interacting particles at low temperatures, and is displayed universally across of athermal solids including crystalline and amorphous structures 
~\cite{o2002random,yoshino2012replica,geng2001footprints}. 
However, in contrast to crystals, amorphous solids form random rigid structures due to the competing interactions between constituent particles, and their static configurations do not correspond to a global energy minimum~\cite{landau2012theory,cui2019theory,biroli2016breakdown}. Although, crystalline and amorphous packings present very different local structure~\cite{phillips1981amorphous}, they display many common elastic properties~\cite{nampoothiri2020emergent,nampoothiri2022tensor}. As the displacement correlations are long-ranged in such systems~\cite{das2021long,das2022displacement}, it is reasonable to question whether the large scale elasticity properties are affected by their microscopic structure~\cite{goldhirsch2002microscopic}. A question of fundamental interest is therefore how the effects of global rigidity are encoded across these disparate networks, and whether such effects can be observed in the fluctuations and correlations in the stress tensor~\cite{vasisht2022residual,barbot2018local,wu2017anomalous}.

Stress correlations provide important information about the collective behavior of disordered systems composed of interacting particles~\cite{lemaitre2014structural,maier2017emergence,klochko2018long,tong2020emergent,rida2022influence,schindler2010numerical,mora2002stress}, and has attracted considerable recent interest~\cite{vinutha2023stress,nampoothiri2022tensor,rida2022influence,vogel2019emergence,dagur2023spatial,PhysRevE.108.015002,mahajan2021emergence}. Consequently analyzing stress correlations can also help better understand the underlying physics of particle packings, such as their degree of rigidity or floppiness, as well as their response to external stimuli such as shear or compression ~\cite{vinutha2023stress,nampoothiri2020emergent}. In this context, the ensemble from which configurations are drawn in order to measure these correlations becomes of central importance.
As the nature of athermal ensembles is at present unclear, with temperature-like variables seeming to govern some of their properties ~\cite{edwards1989theory,blumenfeld2009granular}, 
creating ensembles that are amenable to exact theoretical characterizations becomes important.
In this context, near-crystalline materials that exhibit several properties and interpolate between the well-known physics of crystals and those of amorphous materials~\cite{chaudhuri2005equilibrium,Goodrich2014, tong2015crystals,acharya2021disorder,tsekenis2021jamming,charbonneau2019glassy,mizuno2013elastic}, are useful systems to study the behaviour of athermal ensembles. Recent studies of near-crystalline materials have also revealed several characteristics of fully amorphous solids, including the emergence of quasi-localized modes~\cite{shimada2018spatial,xu2010anharmonic}. The two often separate branches of condensed matter physics, amorphous materials, and crystalline solids are linked through the large-scale elasticity properties displayed in both situations~\cite{otto2003anisotropy,Goodrich2014,gelin2016anomalous}. Gradually introducing disorder into athermal crystalline packings can therefore be used to interpolate between the well-studied physics of crystals and that of amorphous solids.

Several studies on stress correlations in amorphous systems have established an anisotropic $1/r^d$ decay at long distances in $d$ dimensions~\cite{lemaitre2017inherent,lemaitre2018stress,mahajan2021emergence,wu2015anisotropic,chowdhury2016long}.   
These include a variety of contexts such as monodispersed packings quenched from different parent temperatures~\cite{mahajan2021emergence}, isotropic amorphous packings~\cite{henkes2009statistical,lemaitre2018stress,degiuli2018field}, low temperature liquids~\cite{wu2015anisotropic,chowdhury2016long,lemaitre2017inherent} as well as  frictional granular packings~\cite{lemaitre2021stress}.
Field theoretic frameworks have also been developed that predict anisotropic stress correlations in disordered athermal solids~\cite{henkes2009statistical,degiuli2018field,nampoothiri2020emergent,nampoothiri2022tensor}.
Some analytical studies have also suggested a universal  behaviour for stress correlations at large lenthscales~\cite{lemaitre2018stress,lerner2020simple}. However a detailed analysis of this universality across various rigid packings with different interactions as well as a range of microscopic disorder has not been done. In this study, we present exact as well as numerical results for the correlations in the stress tensor, that reveal universal features of stress fluctuations in athermal solids.
We present results for a wide variety of situations, including crystalline, as well as fully amorphous packings. 
We find that the properties of these correlation functions are surprisingly universal across all situations. 
We corroborate this with a microscopic derivation of the correlation functions in near-crystalline configurations as well as numerical results for large disorders, including across an amorphization transition. Our study highlights that the stress correlations in athermal systems at large lengthscale are independent of the microscopic structure and display universal characteristics across various situations, including across a crystalline to amorphous transition.

\section{Microscopic Models and methods}
{Our findings are demonstrated via two paradigmatic model systems for which crystal to amorphous packings and vice-versa can be generated by tuning of certain parameters as detailed below.}

\subsubsection{Repulsive interaction at contact: Harmonic disks}
The first model consists of a system of frictionless disks in two dimensions under varying degrees of overcompression, interacting through a one-sided pairwise potential of the form
\begin{equation}
\begin{aligned}
V_{a_{i j}}\left(\vec{r}_{i j}\right) &=\frac{k}{\alpha}\left(1-\frac{\left|\vec{r}_{i j}\right|}{a_{i j}}\right)^{\alpha},
\end{aligned}
\end{equation}
for $\left|\vec{r}_{i j}\right| / a_{i j}<1$, and $V_{a_{i j}}\left(r_{i j}\right)=0$ for $\left|\vec{r}_{i j}\right| / a_{i j}>1$. Here $\vec{r}_{ij}$ = $\vec{r}_{i}$ - $\vec{r}_{j}$ represents the distance vector between the particles $i$ and $j$, located at positions $\vec{r}_{i}$ and $\vec{r}_{j}$ respectively. 
In this study, we choose $\alpha=2$, to implement a harmonic pairwise potential between particles.
The quenched interaction lengths are expressed as a sum of individual radii as $a_{ij}$ = $a_i$ + $a_j$. 
The tuning of the interaction lengths $a_{ij}$ allows for the transformation from near-crystalline packings to amorphous structures \cite{o2002random,durian1995foam,tong2015crystals,bocquet1992amorphization,mizuno2013elastic}. 
For this model system, we begin with equal sized disks in an overcompressed triangular lattice i.e. a packing fraction ($\phi$) greater than the marginal hexagonal close packing. The lattice constant is given by $R_0 = \sqrt{\frac{\phi}{\phi_c}}$, with $R_0 = 1$ representing the marginal state. The quenched disorder is introduced in the particle radii as
\begin{equation}
 a_i=a_0(1+ \eta \zeta_i).
 \label{ai_harmonic}
\end{equation}
Here ${\zeta}$ represents the quenched disorder in the system and each $\zeta_i$ is  $\pm 1$ (Bidisperse) or varies between $-1/2$ to $1/2$ (Polydisperse). The parameter $\eta$ governs the magnitude of disorder in the system.  Various mechanical properties of this disordered crystal system have been studied in great detail in previous studies~\cite{acharya2020athermal,das2021long,das2022displacement,acharya2021disorder,maharana2022athermal,maharana2022first}.

\subsubsection{Long-ranged attractive interaction: Lennard-Jones}
{The second model that we consider is a system of} particles interacting via a Lennard-Jones (LJ) pairwise potential of the form
\begin{small}
  \begin{equation}
V_{a_{i j}}\left(\vec{r}_{i j}\right) =\epsilon\left[
\left(\frac{a_{i j}}{\left|\vec{r}_{i j}\right|}\right)^{12}-\left(\frac{a_{i j}}{\left|\vec{r}_{i j}\right|}\right)^{6}+\sum_{l=0}^{2}c_{2l}\left(\frac{\left|\vec{r}_{i j}\right|}{a_{i j}}\right)^{2l}
\right],
\end{equation}  
\end{small}
for $\left|\vec{r}_{i j}\right| / a_{i j}<2.5$, and $V_{a_{i j}}\left(r_{i j}\right)=0$ for $\left|\vec{r}_{i j}\right| / a_{i j}>2.5$. The coefficients $c_{2l}$ are chosen to smoothen the potential up to the second order at the cutoff distance.
Here $\epsilon$ sets the microscopic unit of energy and $a_{i j}$ represents the quenched random interaction lengths.  
For this model, the quenched disorder is introduced into half the particles, which are randomly selected and labelled, to be effectively {\it inflated}~\cite{mizuno2013elastic,lerner2022disordered}. 
\begin{equation}
   a_{i j}=\left\{\begin{array}{cc}\lambda_{\mathrm{SS}} & \text { both } i, j \text { are unlabelled } \\ \eta\left(\lambda_{\mathrm{SL}}-\lambda_{\mathrm{SS}}\right)+\lambda_{\mathrm{SS}} & \text { either } i \text { or } j \text { are labelled } \\ \eta\left(\lambda_{\mathrm{LL}}-\lambda_{\mathrm{SS}}\right)+\lambda_{\mathrm{SS}} & \text { both } i \text { and } j \text { are labelled }\end{array}\right.
\end{equation}
It is convenient to introduce a labelling parameter $t_i$ for every particle $i$, with $t_i=0$ if the particle is small and $t_i=1$ if it is large. 
The length parameters $a_{i j}$ are then set as
\begin{equation}
\begin{aligned}
a_{ij}=\lambda_{SS}+&\eta\left[(t_i+t_j)\left(\lambda_{SL}-\lambda_{SS}\right)\right.\\
&\left.+t_i t_j\left(\lambda_{LL}+\lambda_{SS}-2\lambda_{SL}\right)\right].
\end{aligned}
\label{aij_lj}
\end{equation}
Although we may treat the three length parameters $\lambda_{\mathrm{SS}}, \lambda_{\mathrm{LL}}$ and $\lambda_{\mathrm{SL}}$ separately, a particularly simple case is when $\lambda_{\mathrm{SL}} = (\lambda_{\mathrm{SS}}+\lambda_{\mathrm{LL}})/2$, which we focus on. 
Increasing $\eta$ in both models enables us to systematically vary the system between crystalline to amorphous structures. 
Both the harmonic and LJ models display an amorphization transition at $\eta \approx 0.4$ and $\eta \approx 0.6$ respectively.


\subsection{Simulation details}

In order to verify the predictions of our theory
against numerically obtained stress correlations 
in both models, we prepare inherent structure states of the model systems,
for each realization of the quenched disorder. For doing the energy minimization to obtain these states, we use the FIRE energy minimization protocol~\cite{bitzek2006structural}.

To obtain near-crystalline energy minima, we start with a perfect crystal in which we introduce microscopic disorder through the interaction length as given in Eqs.~\eqref{ai_harmonic} and~\eqref{aij_lj}. The energy minimized configuration  obtained this way represents a unique stress balanced state, which we exploit later in order to derive exact results for the correlations between the components of the stress tensor. 

The amorphous packings are obtained using two different approaches. One way is to start from the perfect crystal and gradually increase the strength of the disorder to create amorphous packings without long range order. The second approach is to quench from random initialization. For our simulations, we choose $N=6400$ particles and $N=6498$ particles for commensurate and incommensurate boxes respectively. 

While the  initial packing fraction for the harmonic model is  $\phi=0.92$, we fix the initial pressure to zero using Berendsen barostat~\cite{berendsen1984molecular} for simulating the LJ model. For all the cases, we  perform simulations for disorder strengths  ranging from $\eta=0.001$ to $\eta=0.7$.

 The local stress components for each particle are then computed in the energy minimized configurations. Identifying $\Delta k_x=2\pi/L_x$ and $\Delta k_y=2\pi/L_y$, we perform a discrete Fourier transform of the local stresses. Here, $L_x$ and $L_y$ represent the linear dimensions of the periodic box containing the particles. Then we compute the configurational averaged stress correlations in Fourier space by performing an average of over $400-500$ configurations.

\begin{figure*}[t!]
\centering
\includegraphics[width=0.96\linewidth]{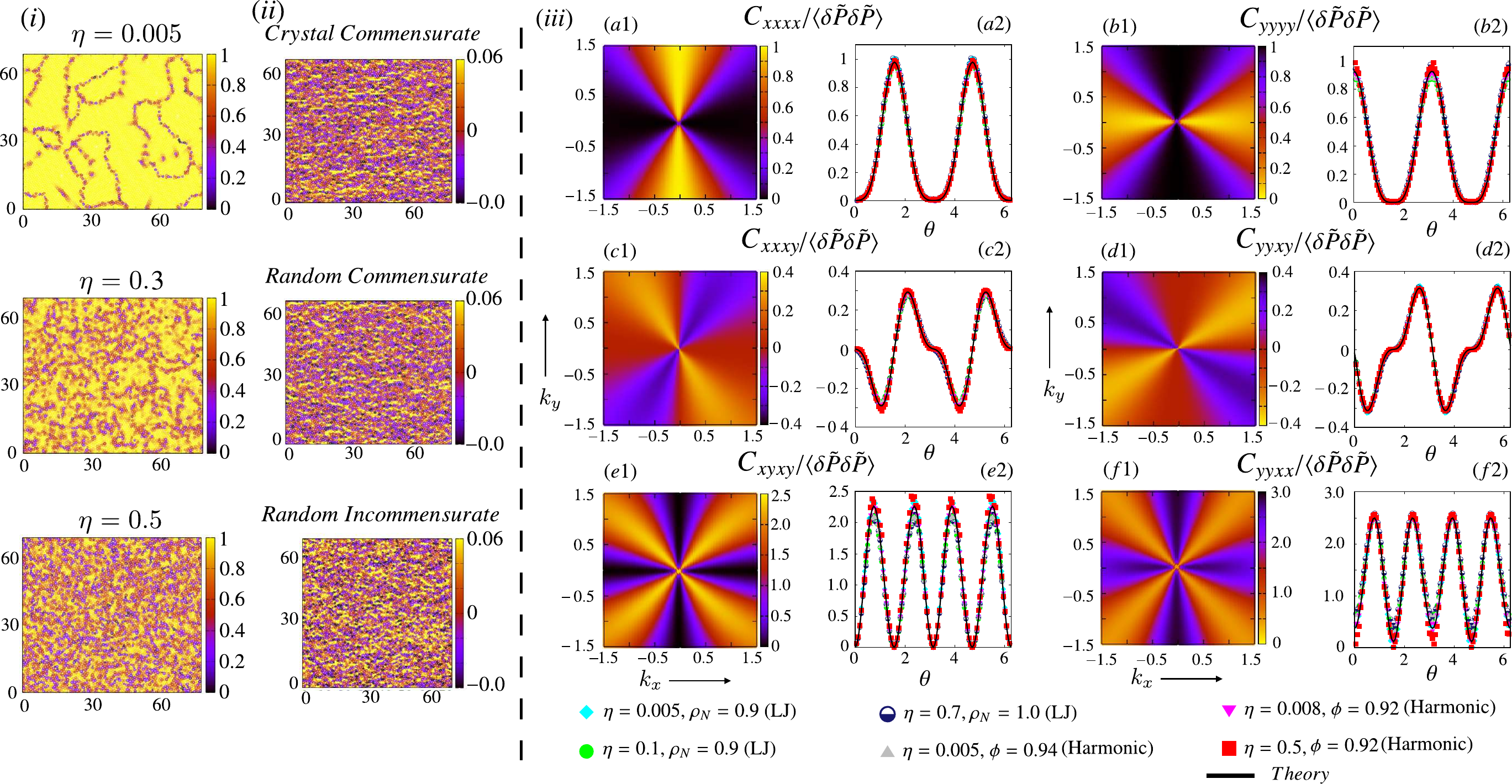}
\caption{
(i) Spatial distribution of the orientational order parameter ($\psi_6$) in the system of harmonic disks for different polydispersities and initial conditions. (ii) Spatial distribution of pressure for different initial conditions, with $\eta = 0.5$; see text for the definition of labels. (iii) $(a1)-(f1)$ Exact predictions for stress correlations in Fourier space using Eq.~\eqref{eq_exact_crystal}. $(a2)$-$(f2)$ Radially integrated stress correlations in Fourier space $C_{ijkl} (\theta)/\langle\delta \tilde{P}\delta \tilde{P}\rangle =\int_{k_{\text{min}}}^{k_{\text{max}}}dk \langle\delta\tilde{\sigma}_{ij}(\vec{k}) \delta\tilde{\sigma}_{kl}(-\vec{k})\rangle/\int_{k_{\text{min}}}^{k_{\text{max}}}dk\langle\delta \tilde{P}(\Vec{k})\delta \tilde{P}(-\Vec{k})\rangle$, with $k_{\text{min}}=0.5$ and $k_{\text{max}}=1.5$; comparison between theory and numerical data. 
The system size is $N=6400$, and the data has been averaged over 400 configurations in order to obtain the ensemble averaged stress-correlations. 
}
\label{fig_corr_crystalline_amorphous}
\end{figure*}
\section{Stress Correlations}

Next, using the energy minimized configurations, we measure the correlations between the components of the stress tensor. The stress field ${\sigma}_{ij}$ within the athermal solid, for a given packing, can be obtained using the particle level force moment tensor~\cite{nampoothiri2020emergent,nampoothiri2022tensor}. 
We measure the fluctuations in the stress field for the grain  situated at position $\vec{r}_g$ as $\delta \sigma_{ij}(\vec{r}_{g}) = \sigma_{ij}(\vec{r}_{g}) - \bar{\sigma}_{ij}$, where $\bar{\sigma}_{ij}$ is the average force moment tensor for a given packing.  
Following recent studies of stress correlations in granular solids and gels~\cite{vinutha2023stress,nampoothiri2020emergent,nampoothiri2022tensor,lois2009stress}, we measure the stress tensor in Fourier space as follows
\begin{equation}
\delta \hat{\sigma}\left(\vec{k}\right)=\sum_{g=1}^{N} \delta  \hat{\sigma}_{g} \exp{\left(i\vec{k}\cdot\vec{r}_{g}\right)}.
\end{equation}
The correlations between the components of the stress tensor in Fourier space can then be obtained as
\begin{eqnarray}
\nonumber
C_{ijkl}(\vec{k}) &=& \langle\delta\tilde{\sigma}_{ij}(\vec{k}) \delta\tilde{\sigma}_{kl}(-\vec{k})\rangle.
\end{eqnarray}
The $\langle \rangle$ represents an average over the different realizations of the microscopic disorder. We perform a disorder average over multiple energy-minimized configurations with the same external conditions: fixed volume and $\eta$ for both models.

The results for stress correlations from our numerical simulations are plotted in Fig.~\ref{fig_corr_crystalline_amorphous}(iii). We find that these correlations in the $k \to 0$ limit, i.e. the large lengthscale limit display surprisingly universal properties.
These are readily apparent from the radially integrated stress correlations in Fourier space $\left(C_{\alpha \beta \gamma \delta} (\theta)=\int_{k_{\text{min}}}^{k_{\text{max}}}dk C_{ijkl}(|\vec{k}|,\theta)\right)$; see Fig.~\ref{fig_corr_crystalline_amorphous}(iii)(b)-(d). 
In order to extract the large lengthscale behaviour and also to avoid effects due to the finite system size, the stress correlations have been integrated in a narrow window in Fourier space with $k_{\text{min}} =  0.5$ and $k_{\text{max}} = 1.5$. This translates to integrating the stress correlations between particles separated less than a distance $L_x/2$ and greater than $4$ particle diameters in real space. Our results are not sensitive to the precise value of $k_{\text{min}}$ (i.e. $k_{\text{min}} \to 0$ gives the same result \ref{sec_kmin_kmax}).

In order to test the universality of our results, we also study systems with different aspect ratios, that allow for commensurate crystalline structures as well as amorphous structures to form. 
Specifically, we construct different initial structures, placing particles with (i) crystalline initial arrangement in a commensurate box (aspect ratio $1:\frac{\sqrt{3}}{2}$) (Crystal Commensurate) (i) random initial points in a commensurate box (Random Commensurate). In order to establish this universal behaviour in amorphous packings that are not associated in a specific way with a
crystalline limit, we also employ (iii) random initial points in an incommensurate box (aspect ratio $1:1$) (Random Incommensurate), for a variety of polydispersities as well as bidisperse packings.
As illustrated in Fig.~\ref{fig_corr_crystalline_amorphous}(i), there is a varying degree of orientational order in the energy minimized states obtained for these different systems, which we quantify using the orientational order parameter $\psi_6 =N^{-1}\sum_{i}\left|z_i^{-1}\sum_{j=1}^{z_i}e^{i6\theta_{ij}}\right|$, where $z_i$ represents the coordination of the $i^{\text{th}}$ particle. 
As illustrated in Fig.~\ref{fig_corr_crystalline_amorphous}(ii), the fluctuations in the pressure display differences at short lengthscales across the different preparation protocols. However, these variations do not affect the large lengthscale behaviour of the stress correlations, as is evident from Fig.~\ref{fig_corr_crystalline_amorphous}(iii), where we have plotted the stress-correlations normalized by the pressure fluctuations showing an excellent collapse. Our numerical tests therefore reveal that the correlations across different situations yield the same angular dependence, with no observable dependence on the degree of orientational order.


\section{Exact Predictions for Near-Crystalline Packings}
As our numerical results reveal the universality of the stress correlations across various situations, it is instructive to derive the exact results which we demonstrate in the near-crystalline limit. For this purpose, it is useful to study the two models with minimal polydispersity added to the quenched interaction lengths between particles. 
In this limit, the uniqueness of the perturbed crystalline state allows us to express the displacements at each site in terms of the underlying quenched disorder $\delta \tilde{r}^{\mu}(\vec{k})=\tilde{G}^{\mu}(\vec{k})\delta \tilde{a} (\vec{k})$, where $\tilde{G}^{\mu}(\vec{k})$ represent the $\mu^{th}$ component of the response Green's functions ~\cite{acharya2020athermal,das2022displacement,acharya2021disorder,das2021long,maharana2022athermal,maharana2022first}, and $\delta \tilde{r}^{\mu}(\vec{k})$ is the Fourier transform of particle displacements in real space i.e.
$\delta \tilde{r}^{\mu}(\vec{k})=\sum_{\vec{r}}e^{i \vec{k}.\vec{r}}\delta r^{\mu}(\vec{r})$.
This formulation allows for analytic computations of the displacement correlations, fluctuations in components of the stress tensor, as well as the interaction energy between stress defects in near-crystalline athermal materials. Taylor expanding the interparticle force about the crystalline positions up to linear order in the displacements $\delta r^{\mu}$, yields $f_{i j}^{\mu}= {f}_j^{\mu(0)} + C_{ j}^{\mu x} \delta x_{i j}+C_{ j}^{\mu y} \delta y_{i j}+C_{ j}^{\mu a} \delta a_{i j} $.
Here $C_j^{\mu \nu}$ are the linear order coefficients of the inter-particle potential at linear order. Here
$\vec{f}^{(0)}$ is the inter-particle force between particles $i$ and $j$, separated by a distance $\Vec{\Delta}_j$ in the initial crystalline arrangement. These coefficients $C_j^{\mu \nu}, \Delta^{\alpha}_{j}$ depend only on the initial crystalline structure, the form of the interaction potential, and the relative position of the neighbour $j$ in the initial crystalline structure. Here $\delta a_{ij}= \delta a_i+\delta a_j$ for the Harmonic model whereas $\delta a_{ij}=\eta (\lambda_{SL}-\lambda_{SS}) (t_i+t_j) $ for the LJ model.
This formulation, developed in the context of harmonic disks \cite{acharya2020athermal,acharya2021disorder}, has now been extended to the LJ system in the near-crystalline limit, with $\delta a_i \equiv t_i$. Next, by imposing the force balance conditions on every grain, the linear order displacement fields can be uniquely obtained. 
Using the exact displacement fields we can express the components of the stress tensor in Fourier space in terms of the microscopic disorder as
\begin{eqnarray}
\delta \tilde{\sigma}_{\alpha \beta}(\vec{k})=S_{\alpha \beta} (\vec{k}) \delta \tilde{a}(\vec{k}),
\label{delsigk}
\end{eqnarray}
where $S_{\alpha \beta} (\vec{k})$ are the relevant source terms in Fourier space that can be derived explicitly from the underlying crystalline lattice and grain disorder (see Supplementary Material for details~\ref{Supplemental_material}).
Explicitly we have
\begin{eqnarray}
\label{eq_Sab}
    &&\hspace{-0.5cm}
    S_{\alpha \beta} (\vec{k})   = \sum_{j}\left[e^{-i \vec{k}.\vec{\Delta}_j}+1\right]C_{j}^{\beta a} \Delta^{\alpha}_{j}+\\
    \nonumber
   &&\sum_{j} \left[e^{-i \vec{k}.\vec{\Delta}_j} -1\right]\left(\sum_{\mu} \Delta^{\alpha}_{j}C_{j}^{\beta \mu}  \tilde{G}^{\mu} (\vec{k})+f_{j}^{\beta (0)}\tilde{G}^{\alpha}(\vec{k})\right),
\end{eqnarray}
where $\mu = x,y$. Eq. \eqref{eq_Sab} has the same form for both models, with the only differences arising in the coefficients and range of interaction. 
For the harmonic model, the range of the interaction determines $z = 6$ neighbours for each particle in the crystalline configuration, as we only need to consider distances up to the first shell, whereas, for the LJ model, the interactions are up to the third shell, with  $z = 18$. The derivation of the exact Green's functions, as well as the correlations between the various stress tensor components, are given in the Supplemental Material~\ref{Supplemental_material}. The underlying microscopic quenched random variables are uncorrelated in real space, which in Fourier space translates to $ \langle\delta \tilde{a}(\vec{k}) \delta \tilde{a}(\vec{k}')\rangle = \frac{\eta^2}{48}\delta_{\vec{k},-\vec{k}'} $ for the harmonic model and $\langle\delta \tilde{a}(\vec{k}) \delta \tilde{a}(\vec{k}')\rangle = \frac{\eta^2}{4}(\lambda_{SL}-\lambda_{SS})\delta_{\vec{k},-\vec{k}'}$ for the LJ model.
Using these microscopic correlations, we arrive at the expressions for the stress correlations in disordered crystals as given in Eqs.~\eqref{eq_exact_crystal} and~\eqref{eq_exact_crystal2}. Specifically, for Harmonic interactions we have
 \begin{equation}
    \begin{aligned}
       \langle \delta \tilde{\sigma}_{\alpha \beta} (\vec{k})\delta \tilde{\sigma}_{\gamma \delta}& (-\vec{k}) \rangle =\frac{\eta^2}{48} S_{\alpha \beta} (\vec{k})S_{\alpha \beta} (-\vec{k}),
       \label{eq_exact_crystal}
    \end{aligned}
\end{equation}
and for the LJ interaction, we have
 \begin{equation}
    \begin{aligned}
       \langle \delta \tilde{\sigma}_{\alpha \beta} (\vec{k})\delta \tilde{\sigma}_{\gamma \delta}& (-\vec{k}) \rangle = \frac{\eta^2(\lambda_{SL}-\lambda_{SS})}{4} S_{\alpha \beta} (\vec{k})S_{\alpha \beta} (-\vec{k}).
       \label{eq_exact_crystal2}
    \end{aligned}
\end{equation}

Stress correlations computed in Fourier space from direct numerics show an exact match with the above predictions, as displayed in Fig.~\ref{fig_corr_crystalline_amorphous}(iii).
Surprisingly, this microscopic theory provides extremely accurate results for a wide range of polydispersities. 
We note that although these configurations have {\it many} broken contacts (particles that have been inter-particle distances further than the interaction range), the theory is able to predict stress correlations with remarkable accuracy. The success of such a microscopic theory in the $k \to 0$ limit stems from the fact that it captures the large length scale behaviour of the system, and therefore is not sensitive to the local structure of the packing.

\section{Crystalline versus Amorphous Packings: comparing stress correlations}

\begin{figure}[t!]
\centering
\includegraphics[scale=0.6]{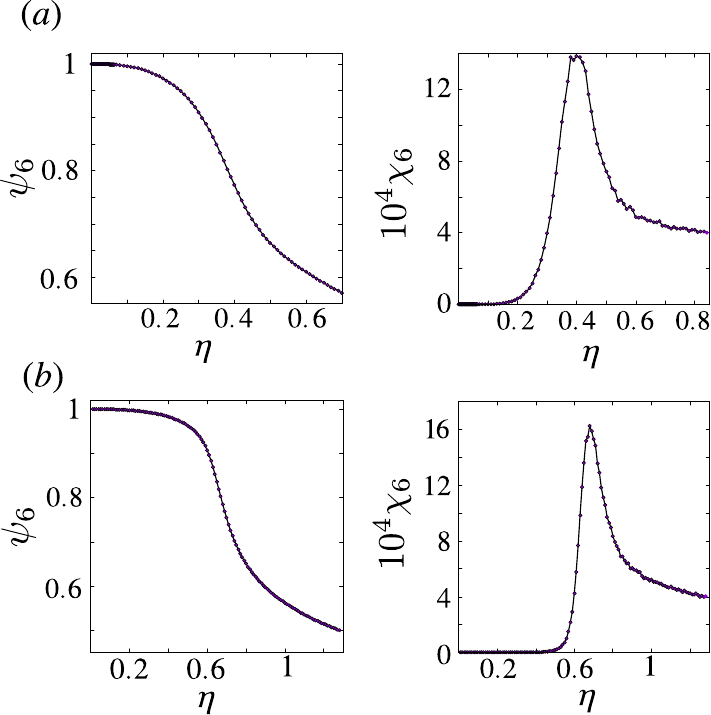}
\caption{Bond orientational order parameter ($\psi_6$) and its fluctuation ($\chi_6$) with variation in the disorder parameter $\eta$ for $(a)$ short ranged Harmonic model and $(b)$ long-ranged LJ model. For each value of $\eta$, $\psi_6,\chi_6$ are obtained by averaging over $50000$ disordered configurations of system size $N=256$.
}
\label{fig_bond_orient}
\end{figure}

 Finally, we investigate the differences between crystalline and amorphous packings through a measurement of the stress correlations.  The transition between these two phases has been studied for both models in many previous studies~\cite{tong2015crystals,mizuno2013elastic}.
The amorphization transition with increasing disorder (polydispersity $\eta$) can be observed via the fluctuation in the bond orientational parameter defined as
\begin{equation}
    \psi_6 = \frac{1}{N}\sum_{i}\left|\frac{1}{z_i}\sum_{j=1}^{z_i}e^{i6\theta_{ij}}\right|.
\end{equation}
Here $z_i$ is the coordination number of the $i^{th}$ particle in the energy minimised configuration and $\theta_{ij}$ is the relative angle made by the bond between particle $i$ and its neighbour $j$ with $x$-axis. For near crystalline systems $\psi_6\to 1 $, whereas in completely amorphous systems $\psi_6$ is typically small.
Further, one can compute the susceptibility of $\psi_6$ as
\begin{equation}
    \chi_6= \langle \psi_6^2\rangle-\langle\psi_6 \rangle^2.
\end{equation}
The susceptibility $\chi_6$  diverges near the amorphization transition, and can therefore distinguish between crystalline and amorphous states.
For the harmonic model, with initial packing fraction $\phi=0.92$, the divergence of $\chi_6$ occurs at a polydispersity of $\eta \approx 0.42$ as plotted in Fig.~\ref{fig_bond_orient}(a) indicating the onset of amorphization transition. For the LJ model, with initial pressure $P=0$ and constant volume, the transition is observed  at $\eta \approx 0.65$ as shown in Fig.~\ref{fig_bond_orient}(b).

\begin{figure}[t!]
\centering
\includegraphics[width=0.9\columnwidth]{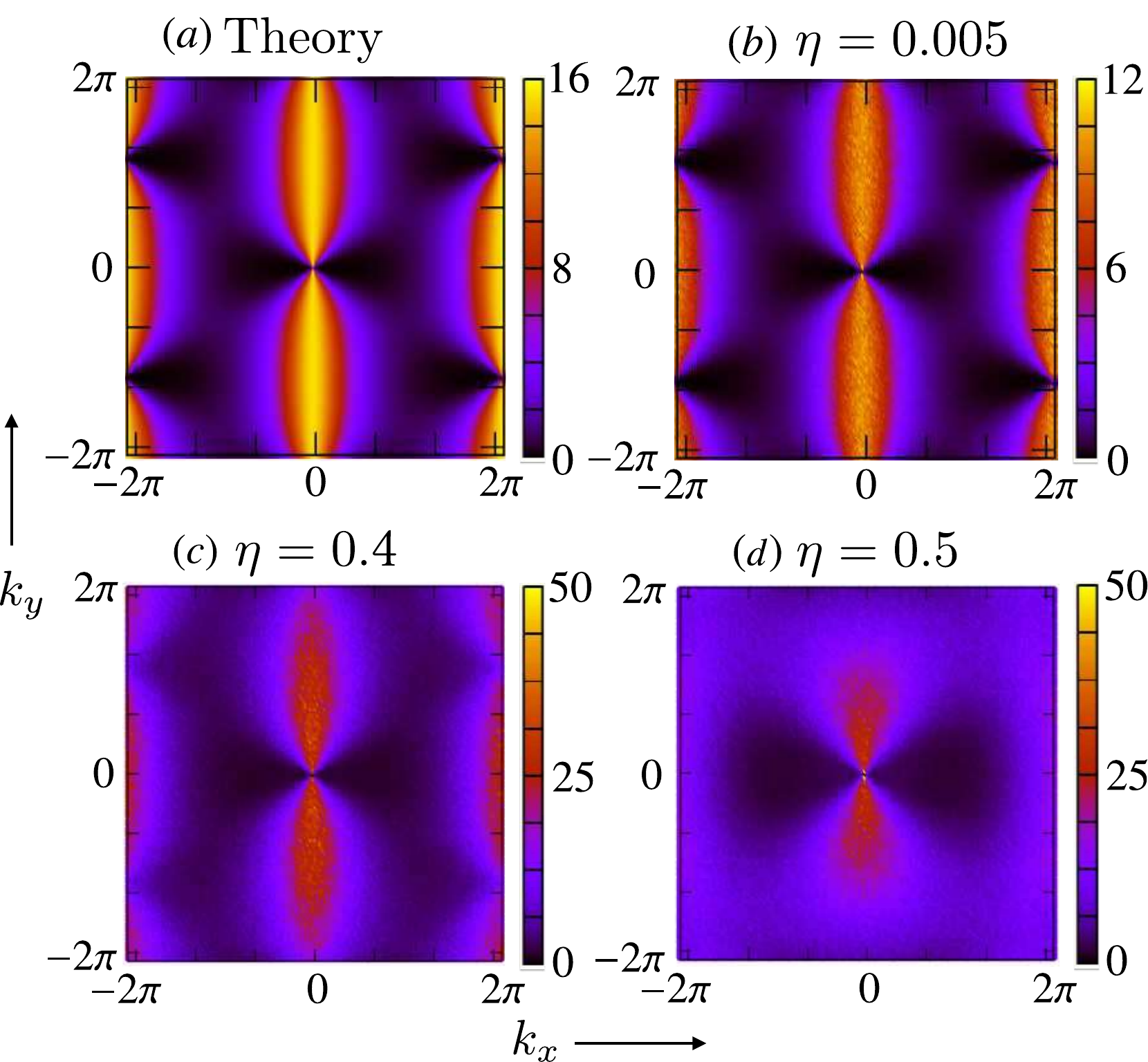}
\caption{Stress correlations ($C_{xxxx}(\Vec{k}) =  \langle \delta \tilde{\sigma}_{xx}(\Vec{k})\delta \tilde{\sigma}_{xx} (-\Vec{k}) \rangle$) in the harmonic model. (a) Theoretical prediction in the $\eta \to 0$ limit, (b)-(d) numerical results with polydispersities varying across the amorphization transition, at fixed initial packing fraction $\phi=0.92$. These stress correlations display the same large lengthscale behaviour, with observable changes at larger values of $|\vec{k}|$, near the edges of the Brillouin zone.
}
\label{fig_stress_correlations_large_polydispersities}
\end{figure}

\begin{figure*}[t!]
\centering
\includegraphics[width=1.60\columnwidth]{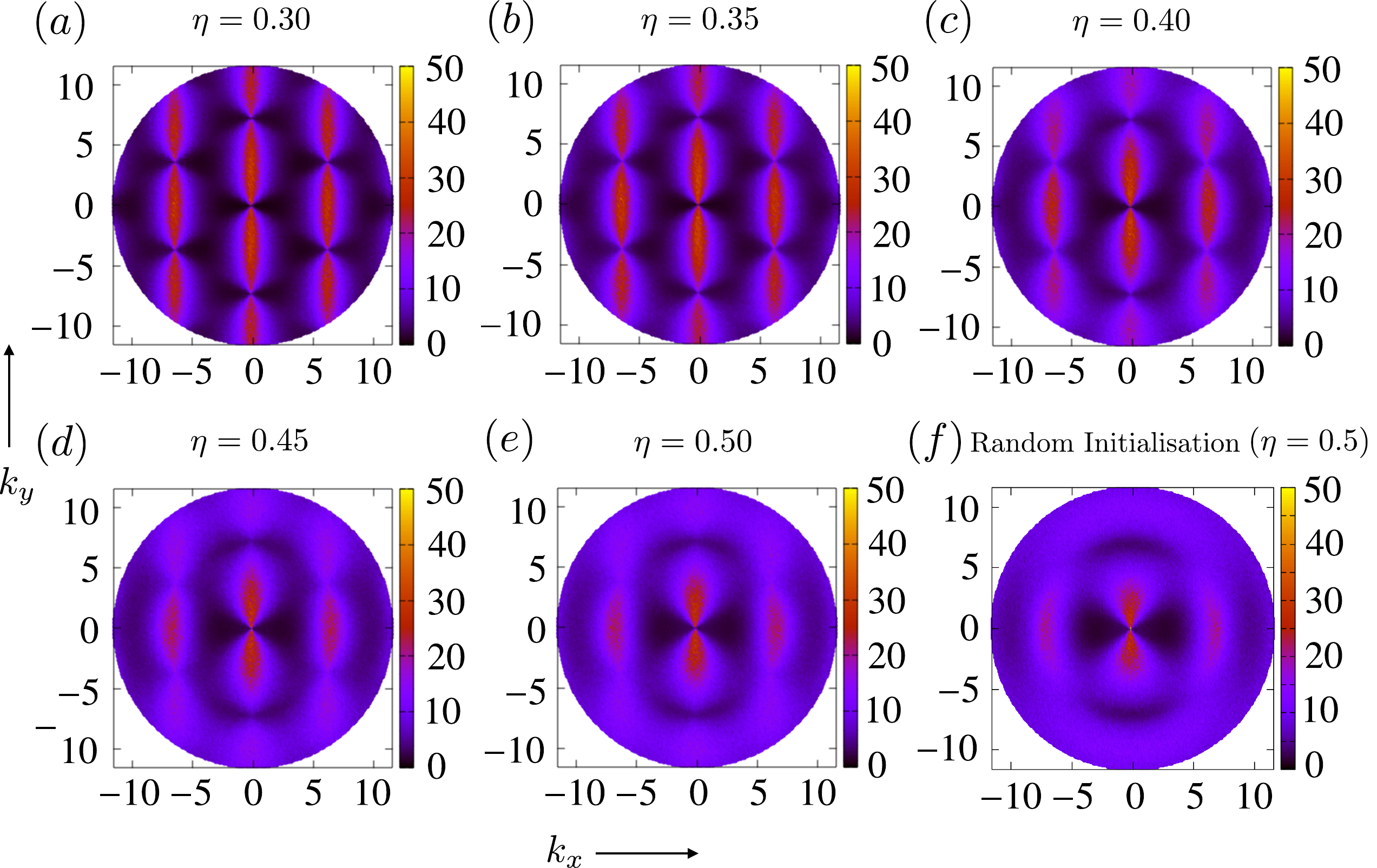}
\caption{Stress correlations ($C_{xxxx}(\Vec{k}) =  \langle \delta \tilde{\sigma}_{xx}(\Vec{k})\delta \tilde{\sigma}_{xx} (-\Vec{k}) \rangle$) in the harmonic model plotted for $|\vec{k}| \le 11.5$. (a)-(e) Numerical results with polydispersities varying across the amorphization transition, at fixed initial packing fraction $\phi=0.92$, (f) stress correlations for jammed packings quenched from random initialization with $\eta=0.5$. These stress correlations display the same large lengthscale behaviour, with observable changes at larger values of $|\vec{k}|$, near the edges of the Brillouin zone.
}
\label{fig_stress_correlations_amorphization}
\end{figure*}

Now we demonstrate how the stress correlations shape up across the amorphization transition. For the harmonic model, we display results  for both near-crystalline and across the amorphization transition ($\eta = 0.4,0.5$) in Fig.~\ref{fig_stress_correlations_large_polydispersities} (b)-(d). These stress correlations do not show any significant change at large lengthscales ($|\vec{k}| \to 0$), as predicted by our exact results for near-crystalline systems.

In the amorphous limit, 
various frameworks have attempted characterization of stress correlations~\cite{henkes2009statistical,degiuli2018field,maier2017emergence,lemaitre2017inherent,lemaitre2018stress,lemaitre2021stress}. In particular, a recently developed ``stress only'' framework to describe stress fluctuations in athermal solids termed Vector Charge Theory of Granular mechanics (VCTG), allows for a computation of the correlations between the components of the stress tensor~\cite{nampoothiri2020emergent, nampoothiri2022tensor}. Within a continuum framework, the local force-balance constraints can be expressed as~\cite{chaikin2000principles,landau2012theory}
$\partial_i \sigma_{ij} = 0$. 
Torque balance leads to a symmetric stress tensor $\sigma_{xy} = \sigma_{yx}$. However, the mechanical force balance is insufficient to solve for all the components of the stress tensor~\cite{ball2002stress}.  
The VCTG framework~\cite{nampoothiri2020emergent}, posits a Gaussian action, with the coupling between the components of the stress tensor representing generalized elastic moduli. This can then be used to compute correlations between the components of the stress tensor that display pinch-point singularities as $|\vec{k}| \to 0$. The predictions from this theory are valid up to lengthscales of the order of the grain diameter: with very short ranged correlations in real space and pinch-point singularities appearing at small $k$~\cite{nampoothiri2020emergent,nampoothiri2022tensor}.
Our microscopic derivations of the stress correlations in the near-crystalline limit (Eq.~\eqref{eq_exact_crystal}) therefore allow for a direct comparison of the stress correlations between an exact theory at low disorder, and the predictions of VCTG. In particular, our microscopic derivation yields an angular dependence for $C_{xyxy} \sim \sin^2(\theta) \cos^2(\theta)$, independent of packing fraction and initial pressure, which matches the predictions from the field theory exactly.

Further, in the near-crystalline limit, our theory provides the exact expressions for the correlations in the entire range of $k$ and not just in the small $k$ limit. Being derived from a reference crystalline structure, these exact predictions are periodic in Fourier space, with the Brillouin zone being determined by the periodicity of the crystal.
We therefore expect the symmetry of the crystal to be present in the correlations at shorter lengthscales. 
The match with the near-crystalline predictions is expected to get worse near the Brillouin zone edges as disorder is increased. This is illustrated in Fig.~\ref{fig_stress_correlations_amorphization}, where we display the changes in the correlations $C_{xxxx}$ with varying polydispersity for the Harmonic model across the amorphization transition. From Fig.~\ref{fig_bond_orient}(a) we observe the amorphization transition occur about $\eta \approx 0.42$. We also observe that the periodic structure of stress correlations in Fourier space starts to vanish after the transition. This represents an intriguing signature of the well-studied crystal to amorphous transition and provides a direct order parameter with measurable differences. The structure of the correlations in Fourier space warrants further investigation.

\section{conclusion and discussion}

In this paper, we have demonstrated, both numerically as well as theoretically, the universality of stress correlations in static athermal solids. Our results for near-crystalline, as well as amorphous packings, demonstrate that although real-space measures of orientational order vary in these structures, the correlations in the {\it fluctuations} of the stress tensor at large lengthscales remain unaffected, as suggested in previous studies~\cite{lemaitre2018stress,lerner2020simple}.
This allows for correlations extracted from a reference crystalline configuration to correctly predict stress correlations even in amorphous packings.

Several intriguing directions for further research remain. The universality of our findings has implications for diverse systems such as granular and glassy materials, as well as biological tissues. It would be interesting to analyze in detail the features of the crystal to amorphous transition using stress correlations, to test whether a stress-only order parameter is able to capture the significant features of the transition between the phases. Another aspect that remains to be probed is the dependence of the correlations on the pre-stress of the system, which can play an important role in the nature of the vibrational eigenmodes and consequently in the stability of amorphous solids~\cite{zhang2022prestressed,shimada2018spatial}.
As our microscopic theory correctly predicts the large lengthscale behaviour, it would be useful to derive a Lagrangian that incorporates the effects of the microscopic disorder, which could help explain features in the correlations observed from a coarse grained approach~\cite{nampoothiri2020emergent}. 
Finally, although the static limit reveals a striking universality between crystalline and amorphous structures, it would be very interesting to study dynamical signatures through stress correlations, which can reveal differences between the well-studied elastodynamics of crystals~\cite{ostoja2019ignaczak} and the dynamics of amorphous systems.

\section{Acknowledgments}
We thank Surajit Chakraborty, Bulbul Chakraborty, Subhro Bhattacharjee, Chandan Dasgupta, and Edan Lerner for useful discussions. This project was funded by intramural funds at TIFR Hyderabad from the Department of Atomic Energy (DAE), Government of India. The work of K.~R. was partially supported by the SERB-MATRICS grant MTR/2022/000966.

\bibliography{Stress_Correlations_Bibliography}

\clearpage

\begin{widetext}

\section*{\large Supplemental Material for\\ ``Universal stress correlations in crystalline and amorphous packings"}
\label{Supplemental_material}

In this document, we provide supplemental figures and details related to the results presented in the main text. We provide details of the derivations of stress correlations in near-crystalline packings, as well as detailed comparisons of analytic results with direct numerical simulations. We also provide additional numerical evidence for the universal behaviour of stress correlations at large length scales.

\maketitle



\section{Exact results for near-crystalline systems}
Following previous studies of near-crystalline systems~\cite{acharya2020athermal,das2021long,das2022displacement,acharya2021disorder,maharana2022athermal,maharana2022first}, linear order displacement fields can be expressed in Fourier space as
\begin{equation}
\begin{aligned}
    \delta \tilde{r}^{\mu} (\Vec{k}) = 
     &\left(\sum_{\nu}( A^{-1})^{\mu \nu}(\vec{k})D^{\nu}(\vec{k})\right)\delta\Tilde{a}(\Vec{k}),
\end{aligned}   
\label{disp_sr}
\end{equation}
where $\delta \tilde{r}^{\mu} (\Vec{k})$ and $\delta \tilde{a} (\vec{k})$ correspond to the discrete Fourier transform of $\delta r^{\mu}_i$ and $\delta a_i$ in real space respectively. Here $\delta\Tilde{a}(\Vec{k})= \eta a_0 \tilde{\zeta}(\vec{k})$ for the Harmonic model and $\delta\Tilde{a}(\Vec{k})$ is replaced by $\eta(\lambda_{SL}-\lambda_{SS}) \tilde{t}(\vec{k})$ for the LJ model. Additionally, we have
\begin{equation}
\begin{aligned}
\hspace{1cm}A^{\mu \nu}(\vec{k})=\sum_{j} \left(1-e^{-i\vec{k}.\vec{\Delta_j}}\right) C_{i j}^{\mu \nu},&\hspace{0.3cm}
   D^{\nu}(\vec{k})=
    \sum_{j}\left(1+e^{-i\vec{k}.\vec{\Delta}_{j}}\right)C^{\nu a}_{ij}.
\end{aligned}    
\label{disp_K}
\end{equation}
Here $C_{ij}^{\alpha \beta}$ are the linear order Taylor coefficients that appear in the expansion of the $\alpha$-component of the interaction force between particles $i$ and neighbouring particles $j$. $\vec{\Delta}_j$ are displacement vectors between particles $i$ and $j$ in the initial crystalline arrangement. For Lennard-Jones interactions, we choose $\lambda_{SL}=(\lambda_{SS}+\lambda_{LL})/2$. Next, the displacements of particles from their crystalline positions can be obtained by taking an inverse Fourier transform of Eq.~\eqref{disp_K} as
\begin{equation}
    \delta r^{\mu} (\Vec{r}) = \frac{1}{N}\sum_{\Vec{k}}\exp{(-i \Vec{k}.\Vec{r})} \delta \tilde{r}^{\mu}(\Vec{k}).
    \label{disp}
\end{equation}

For the case of the minimally polydisperse system, it is possible to derive the stress correlations using a perturbation theory about the crystalline ordered state. The components of the macroscopic stress tensor can be expressed as
\begin{equation}
\begin{aligned}
     \Sigma_{\alpha \beta} &=V^{-1}\sum_{\langle i j \rangle} r_{ij}^{\alpha}f_{ij}^{\beta}\\
       &=V^{-1}\sum_{\langle i j \rangle}\left(r^{\alpha(0)}_{ij}f_{ij}^{\beta (0)}+r^{\alpha(0)}_{ij}\delta f_{ij}^{\beta }+\delta r^{\alpha}_{ij}f_{ij}^{\beta (0)}+\delta r^{\alpha}_{ij} \delta f_{ij}^{\beta }\right).
\end{aligned}
\end{equation}
where $r_{ij}^{\mu(0)}$ and $f_{ij}^{\mu(0)}$ represent the $\mu = x,y$ component of the interparticle separation and force between particle $i$ and $j$ in the crystalline arrangement. 

For small polydispersity both the change in inter-particle forces as well as the deviation of the particle positions ($\delta r$) from their crystalline positions will be of order $\delta a \sim\eta$, and we consider only the linear order terms in these expressions. To linear order, the change in the stress tensor can be expressed as
\begin{equation}
\begin{aligned}
    \delta \Sigma_{\alpha \beta}=& \Sigma_{\alpha \beta} - V^{-1}\left(\sum_{\langle i j \rangle}r^{\alpha(0)}_{ij}f_{ij}^{\beta (0)}\right) = V^{-1}\sum_{\langle i j \rangle}(r^{\alpha(0)}_{ij}\delta f_{ij}^{\beta }+\delta r^{\alpha}_{ij}f_{ij}^{\beta (0)})\\
    \delta \Sigma_{\alpha \beta}=& \sum_i \left(V^{-1}\sum_j (r^{\alpha(0)}_{ij}\delta f_{ij}^{\beta }+\delta r^{\alpha}_{ij}f_{ij}^{\beta (0)})\right)= V^{-1}\sum_i \delta\sigma_{\alpha \beta}(\Vec{r}_i),
\end{aligned}
\end{equation}
where the local stress can be written as
\begin{equation}
\begin{aligned}
    \delta \sigma_{\alpha \beta}(\Vec{r}_i)=&\sum_j (r^{\alpha(0)}_{ij}\delta f_{ij}^{\beta }+\delta r^{\alpha}_{ij}f_{ij}^{\beta (0)})
=\sum_{j}\left[\Delta_{j}^{\alpha }\sum_{\nu}C_{ij}^{\beta \nu} \delta r_{ij}^{\nu}+\Delta_{j}^{\alpha }C_{ij}^{\beta a} \delta a_{ij}+\delta \Delta^{\alpha}_{j}f_{ij}^{\beta (0)}\right].
\end{aligned}   
\end{equation}

\begin{figure}[h!]
\centering
\includegraphics[scale=0.30]{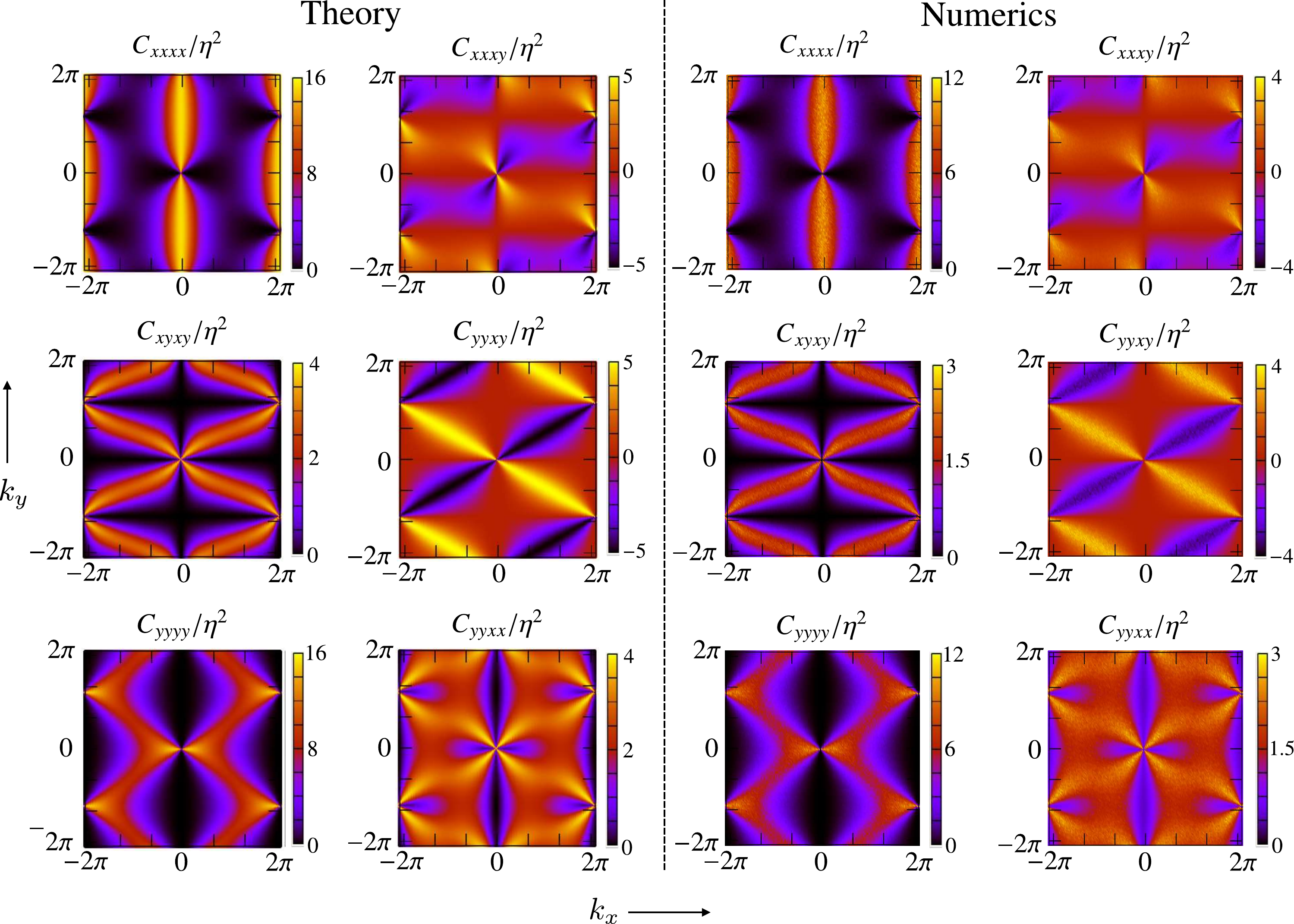}\caption{Theoretical predictions in Fourier space for the correlations between various components of the stress tensor, for the near-crystalline system. (Left) Theoretically obtained values for stress correlation (Right) numerical results for small values of polydispersity (i.e. $\eta=0.005$) for harmonic disks. 
}
\label{fig_theoretical_kspace}
\end{figure}

 Next, taking a Fourier transform of the local stress components, $\delta \sigma_{\alpha \beta}(\vec{r}_i)$, we arrive at the stress tensor in Fourier space
\begin{equation}
\begin{aligned}
    \delta \tilde{\sigma}_{\alpha \beta}(\vec{k})=&\sum_{i} e^{i \vec{r}_i. \vec{k}} \delta \sigma_{\alpha \beta}(\vec{r}_i)
    \\   =&\sum_{j}\left[\Delta_{j}^{\alpha }\sum_{\nu}C_{j}^{\beta \nu} \left[-1+e^{-i \vec{k}.\vec{\Delta}_j}\right]\tilde{\delta r}^{\nu}(\vec{k})+\Delta_{j}^{\alpha }C_{j}^{\beta a}\left[1+e^{-i \vec{k}.\vec{\Delta}_j}\right] \tilde{\delta a}(\vec{k})+\left[-1+e^{-i \vec{k}.\vec{\Delta}_j}\right]\tilde{\delta r}^{\alpha}(\vec{k})f_{j}^{\beta (0)}\right]\\
    =&S_{\alpha \beta} (\vec{k}) \delta \tilde{a}(\vec{k}),
\end{aligned}    
\label{delsig_k}
\end{equation}
The source term therefore has the explicit form
given in Eq.~\eqref{eq_Sab} in the main text. 
The correlations between the quenched microscopic disorder is given by
\begin{equation}
\begin{aligned}
\langle\delta \tilde{a}(\vec{k}).\delta \tilde{a}(\vec{k}')\rangle =& \frac{\eta^2}{48}\delta_{\vec{k},-\vec{k}'},\hspace{1.5cm}\text{(Harmonic)},  \\
\langle\tilde{t}(\vec{k}).\tilde{t}(\vec{k}')\rangle=&\frac{1}{4}\delta_{\vec{k},-\vec{k}'},\hspace{1.7cm}\text{(LJ)} .
\end{aligned}
\end{equation}
Therefore the correlation in the components of the stress tensor in Fourier space $ \delta \tilde{\sigma}_{\alpha \beta}(\vec{k})$ can be expressed as

\begin{equation}
\begin{aligned}
   C_{\alpha \beta \gamma \delta} (|\vec{k}|,\theta)=\langle \delta \tilde{\sigma}_{\alpha \beta}(\vec{k}). \delta \tilde{\sigma}_{\gamma \delta}(\vec{k}')\rangle &= S_{\alpha \beta} (\vec{k})S_{\gamma \delta} (\vec{k}')\langle\delta \tilde{a}(\vec{k}).\delta \tilde{a}(\vec{k}')\rangle\\
    &=\frac{N\eta^2}{48}\delta_{\vec{k},-\vec{k}'}S_{\alpha \beta} (\vec{k})S_{\gamma \delta} (\vec{k}'),\hspace{2.5cm}\text{(Harmonic)}\\
    &=\frac{N\eta^2(\lambda_{SL}-\lambda_{SS})^2}{4}\delta_{\vec{k},-\vec{k}'}S_{\alpha \beta} (\vec{k})S_{\gamma \delta} (\vec{k}'),\hspace{0.5cm}\text{(LJ)}.
\end{aligned}  
\label{stress_correlation_k}
\end{equation}
which is Eq.~\eqref{eq_exact_crystal} in the main text. We also define $C_{\alpha \beta \gamma \delta} (\theta)  =\int_{k_{\text{min}}}^{k_{\text{max}}}dk \langle \delta \tilde{\sigma}_{\alpha \beta}(\vec{k}). \delta \tilde{\sigma}_{\gamma \delta}(\vec{k}')\rangle $.

\begin{figure*}[t!]
\centering
\includegraphics[width=1.0\linewidth]{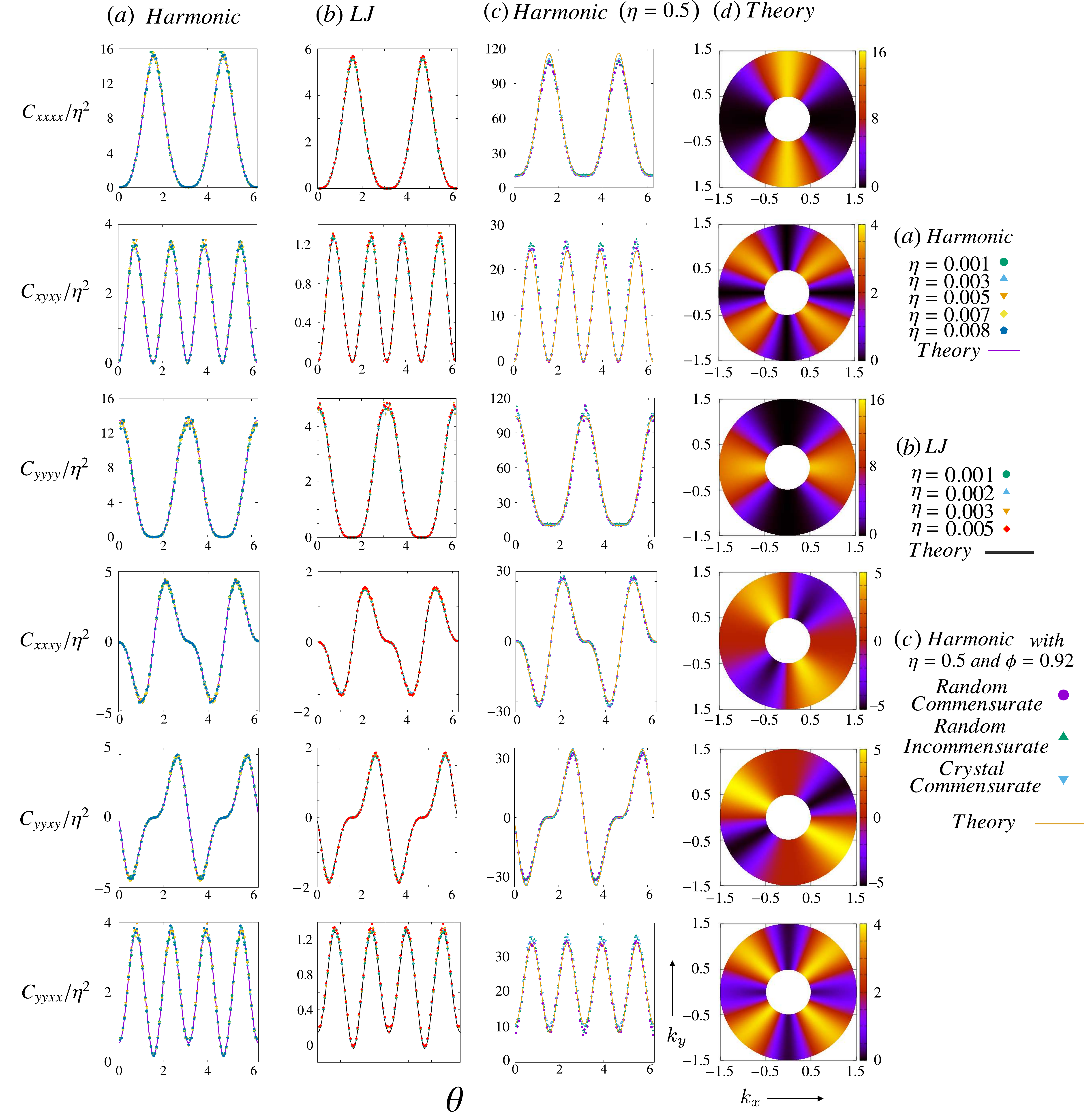}
\caption{
Radially integrated stress correlations in Fourier space $\left(C_{\alpha \beta \gamma \delta} (\theta)/\langle\delta a^2 \rangle  =\int_{k_{\text{min}}}^{k_{\text{max}}}dk S_{\alpha \beta}(k,\theta)S_{\gamma \delta} (-k,\theta) \right)$, with $k_{\text{min}}=0.5$ and $k_{\text{max}}=1.5$. $(a)$ Stress correlations in a system of $N=6400$ harmonic disks, in a fixed volume with initial packing fraction $\phi=0.92$, with polydispersity $\eta$ varying from $0.001$ to $0.005$. The data has been averaged over 400 configurations. $(b)$ Stress correlations in a system of $N=6400$ particles interacting via Lennard Jones potentials, with zero pressure and polydispersity $\eta$ varying from $0.005$ to $0.100$. $(c)$  Stress correlations in amorphous structures of harmonic disks, with polydispersity $\eta=0.5$ and fixed volume and initial packing fraction $\phi=0.92$ for various initialization conditions prior to energy minimization: (i) placing particles randomly in a commensurate box, (ii) placing particles randomly in an incommensurate box, (iii) starting with a crystalline arrangement of particles in a commensurate box. $(d)$ Theoretical predictions for stress correlations $C_{\alpha \beta \gamma \delta}/\eta^2$ in $k_x-k_y$ plane plotted in the range $0.5<|\Vec{k}|<1.5$, which is used in our radial integration in $(a)-(c)$.}
\label{fig_corr_crystalline_amorphous_supp}
\end{figure*}

\begin{figure*}[h!]
\centering
\includegraphics[width=0.90\linewidth]{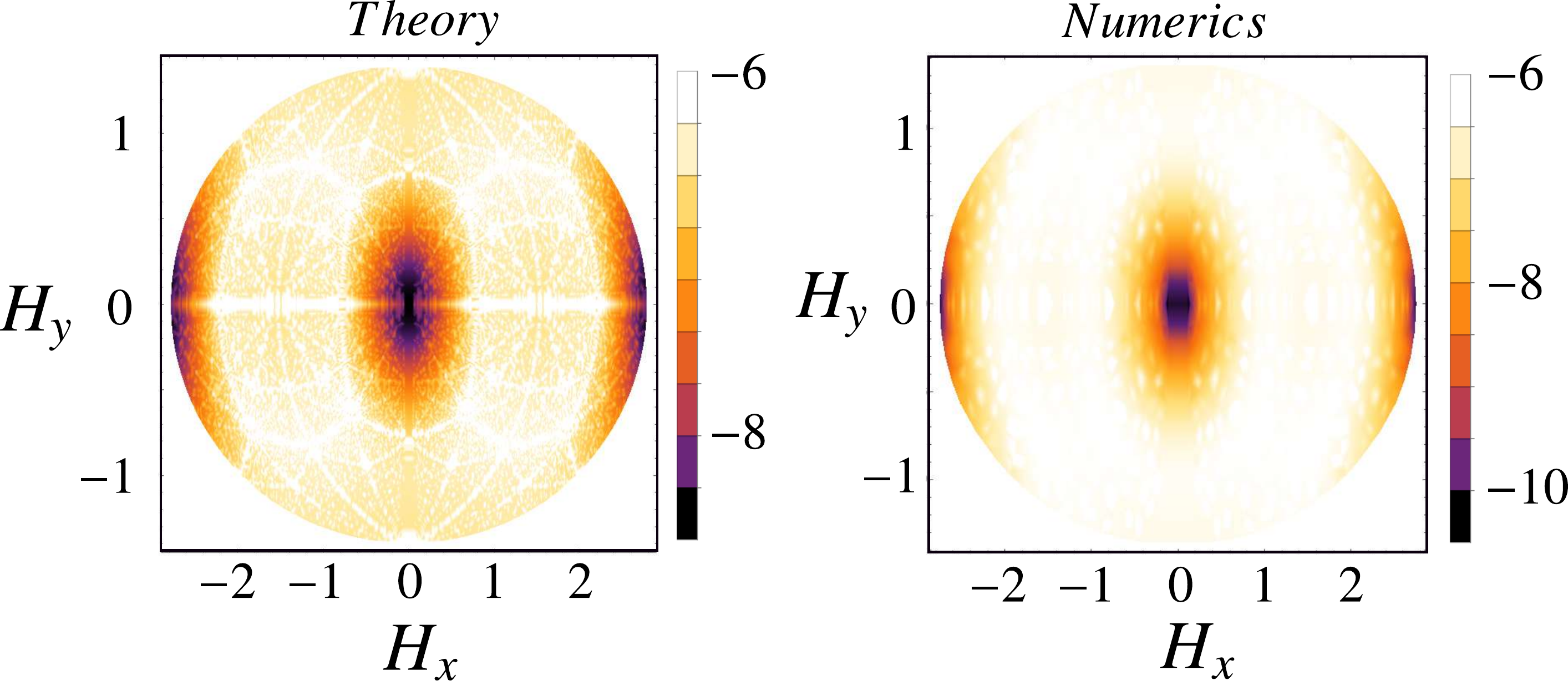}
\caption{ 
Correlations in the local stresses ($C_{xxxx}(\theta,\phi)$) in Fourier space in energy minimized disordered crystals in three dimensional disordered crystals. These have been plotted using a Hammer projection as detailed in Eq.~\eqref{Hammer}. Here we have only plotted for $(H_x/2\sqrt{2})^2+(H_y/\sqrt{2})^2 \le 1$. The above correlations are for the near-crystalline Harmonic system of size $N=108000$ with a polydispersity of $\eta=0.001$.}
\label{fig_corr_3d}
\end{figure*}

\section{Stress correlations in three dimensional systems}
\label{sec_3d}
The theoretical method developed for the change in local stress fields and displacement fields in the main text can also be extended to the case of three dimensional near-crystalline materials. For this case, the number of nearest neighbours is $16$ for an FCC arrangement of particles. So for both the Harmonic and LJ model, the Fourier transform of change in local stress components can still be written as $\delta \tilde{r}^{\alpha }(\vec{k})=G^{\alpha} (\vec{k}) \delta \tilde{a}(\vec{k})$ and $\delta \tilde{\sigma}_{\alpha \beta}(\vec{k})=S_{\alpha \beta} (\vec{k}) \delta \tilde{a}(\vec{k})$, where $\Vec{k}$ corresponds to reciprocal lattice vectors for an FCC lattice. Here for the 3D Harmonic model, $S_{\alpha\beta}$ in $|\vec{k}|\to0$  limit has the following form:
\begin{equation}
   \lim_{|\vec{k}|\to0} S_{\alpha \beta}(|\vec{k}|,\theta,\phi)=\frac{g_{\alpha\beta} (\theta,\phi)}{h (\theta,\phi)}.
\end{equation}

The correlations between different components of stress in ${|\Vec{k}|\to0}$ limit can be expressed as
\begin{equation}
    \begin{aligned}
     C_{\alpha \beta \gamma\delta}(\theta,\phi)=&\lim_{|\Vec{k}|\to0}\left< \delta \tilde{\sigma}_{\alpha \beta}(\vec{k}). \delta \tilde{\sigma}_{\gamma \delta}(-\vec{k})\right> \\
      =&\lim_{|\Vec{k}|\to0}\left[\langle \delta \tilde{a} (\vec{k})\delta \tilde{a} (-\vec{k})\rangle S_{\alpha \beta} (\vec{k})S_{\gamma \delta} (-\vec{k})\right] = \frac{\eta^2}{48}\frac{g_{\alpha \beta} (\theta,\phi)g_{\gamma \delta} (\theta,\phi)}{h(\theta,\phi)^2}.
       \label{eq_exact_crystal_3d}
    \end{aligned}
\end{equation}

Instead of representing $C_{\alpha \beta \gamma \delta}$ as a function of spherical coordinates $(\theta,\phi)$ we can use Hammer projection for its representation given as,
\begin{equation}
    \begin{aligned}
        H_{x}=&\frac{2\sqrt{2}\cos{\left(\theta-\pi/2\right)}\sin{\left(\phi/2\right)}}{\sqrt{1+\cos{\left(\theta-\pi/2\right)}\cos{\left(\phi/2\right)}}},\\
        H_{y}=&\frac{\sqrt{2}\sin{\left(\theta-\pi/2\right)}}{\sqrt{1+\cos{\left(\theta-\pi/2\right)}\cos{\left(\phi/2\right)}}}.
    \end{aligned}
    \label{Hammer}
\end{equation}

These correlations at large lengthscales are displayed in Fig.~\ref{fig_corr_3d}, and are consistent with observations in other amorphous systems~\cite{nampoothiri2020emergent,nampoothiri2022tensor,vinutha2023stress}.

\section{Stress correlations in real space}
\label{sec_real_space_correlation}
The stress correlations in real space can be represented using the correlations in Fourier space as
\begin{equation}
    \begin{aligned}
        \left< \delta \sigma_{\alpha \beta}(\vec{r}). \delta \sigma_{\gamma \delta}(\vec{r}')\right>=\frac{1}{N}\sum_{\vec{k}}\left< \delta \tilde{\sigma}_{\alpha \beta}(\vec{k}). \delta \tilde{\sigma}_{\gamma \delta}(-\vec{k})\right>e^{i \Vec{k}.(\vec{r}-\vec{r}')}.
    \end{aligned}
\end{equation}
The correlations computed in real space are displayed in Fig.~\ref{fig_corr_real}, and display very short ranged behaviour.

\begin{figure*}[h!]
\centering
\includegraphics[width=0.90\linewidth]{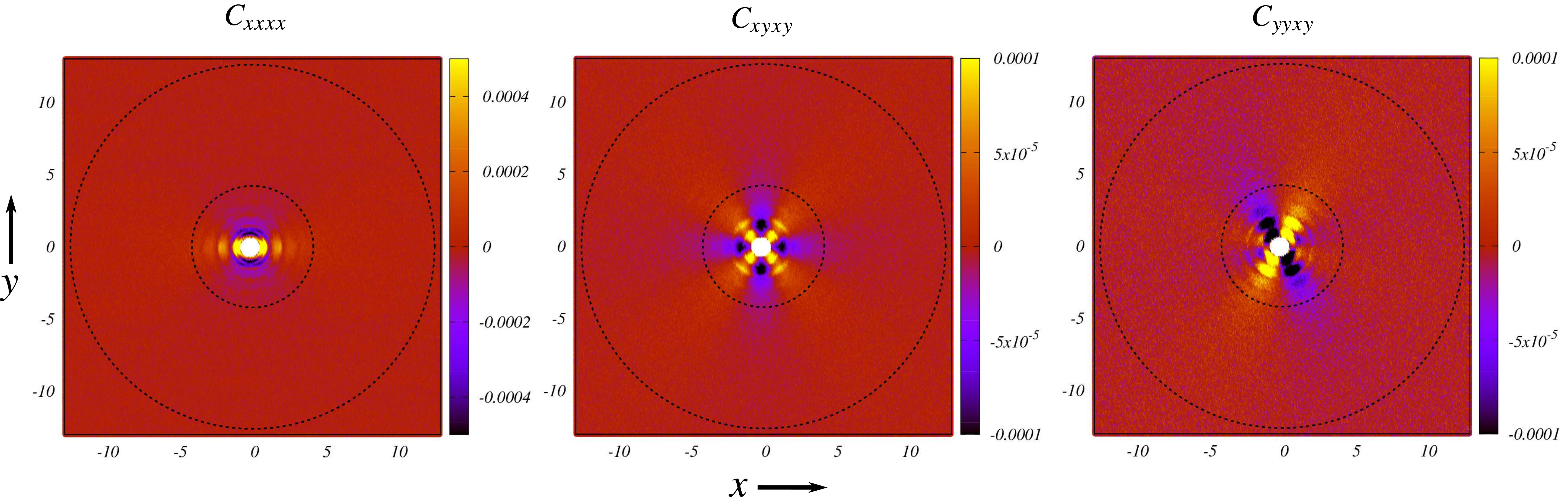}
\caption{
Correlations of the stress tensor in real space for the Harmonic model. Here the two dotted circles correspond to two different values of $|\vec{k}|$ i.e the inner circle of radius $4.2$ corresponds to $k_{\text{max}}=1.5$ whereas the outer circle of radius $12$ corresponds to $k_{\text{min}}=0.5$. We note that although long-ranged power-law correlations are present, the stress correlations in real space decay very quickly at the granular lengthscale.
}
\label{fig_corr_real}
\end{figure*}
\section{Effect of coarse graining}
\label{sec_kmin_kmax}
In order to test the effects of changing the coarse graining lengthscale on the stress correlations, we have varied both the integration limits $k_{\text{min}}$ and $k_{\text{max}}$ with little to no differences in the measured correlations. For different values of $k_{\text{max}}$ we have obtained similar results, as detailed in Figure~\ref{fig_stress_corr_k_vs_kmax}. We find that the stress-correlations are completely independent of the value of $k_{\text{min}}$ for the ranges we have chosen.

\begin{figure}[h!]
\centering
\includegraphics[scale=0.75]{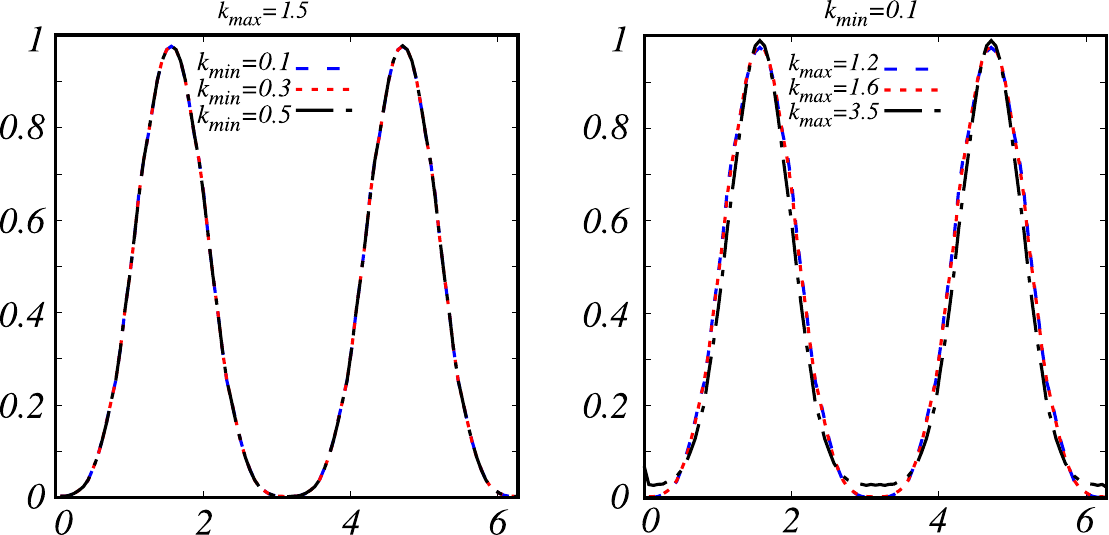}
\caption{
(a) $C_{xxxx}$ for different values of $k_{\text{min}}$ for a fixed $k_{max}=1.5$. The angular behaviour of stress-correlations does not change as we decrease $k_{\text{min}}$. (b) $C_{xxxx}$ for different values of $k_{\text{max}}$ for a fixed $k_{min}=0.1$. The angular behaviour of the stress correlations does not change for $k_{max}=1.2,1.6$, but shows a small variation for $k_{max}=3.5$.
}
\label{fig_stress_corr_k_vs_kmax}
\end{figure}


\newpage


\clearpage
\end{widetext}

\end{document}